\documentstyle[12pt,infnprep,epsfig]{article}
\newcommand{\Header}{
  \begin{tabular}{rl}
  \hspace{-.4cm}\includegraphics{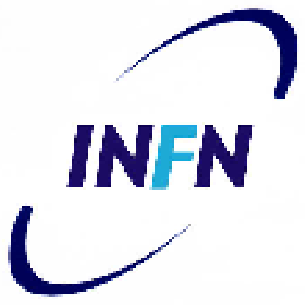} &
    \renewcommand{\arraystretch}{0.5}
    \begin{tabular}{r}
      {\hspace{1cm}~\LARGE\sffamily ISTITUTO NAZIONALE DI FISICA NUCLEARE}\\
      \\
    \end{tabular}
  \end{tabular}
\begin{center}
      {\Large\sffamily Sezione di Trieste}\\
\end{center}
    \renewcommand{\arraystretch}{1}
\vskip 0.5cm
\rule{16.0cm}{0.09mm}
\vskip 1.5cm
  \begin{flushright}
      {\underline{\bf INFN/AE-05/03}}\\    
      {\bf 9 Maggio 2005} \\      
  \end{flushright}
}
\begin{document}
\begin{titlepage}
\title
  {\Header \large \bf ELECTROWEAK CORRECTIONS UNCERTAINTY ON THE $\mathbf{W}$ MASS MEASUREMENT AT LEP}
\author{
   Fabio Cossutti \\
{\it INFN, Sezione di Trieste, I-34127 Trieste, Italy} 
} 
\maketitle
\baselineskip=14pt

\begin{abstract}
The systematic uncertainty on the $W$ mass and width measurement
resulting from the imperfect knowledge of electroweak radiative
corrections is discussed. The intrinsic uncertainty in the 4-$f$
generator used by the DELPHI Collaboration is studied following the
guidelines of the authors of {\tt YFSWW}, on which its radiative
corrections part is based.  The full DELPHI simulation, reconstruction
and analysis chain is used for the uncertainty assessment. A
comparison with the other available 4-$f$ calculation implementing DPA
$\mathcal{O}(\alpha)$ corrections, {\tt RacoonWW}, is also presented.
The uncertainty on the $W$ mass is found to be below 10 MeV for all
the $WW$ decay channels used in the measurement.
\end{abstract}

\vspace*{\stretch{2}}
\begin{flushleft}
\end{flushleft}
\begin{flushright}
  \vskip 1cm
\small\it Published by {\bf SIS-Pubblicazioni}\\
Laboratori Nazionali di Frascati
\end{flushright}
\end{titlepage}
\pagestyle{plain}
\setcounter{page}2
\baselineskip=17pt
\section{Introduction}

Precision tests of the Standard Model in the $W$ sector have been one
of the main issues of the LEP2 physics program. In this context the
measurement of the $W$ mass is one of the most interesting tests. Due
to the high precision which is experimentally achievable, about 0.05\%
in the LEP combination, it is important to have a robust estimate of
all the possible systematic uncertainties.

Electroweak radiative corrections on $WW$ events, which are used for
the $W$ mass and width measurements, and more generally on 4-$f$
events, have been an important issue since LEP2 beginning. After the
LEP2 Workshop of 1995~\cite{lep2} it has been clear that the simple
radiative corrections approach based on the Improved Born
Approximation (IBA) is not sufficient to obtain a theoretical
precision smaller than the experimental foreseen one in precise $W$
physics measurements.

At the 2000 LEP2 Monte Carlo workshop~\cite{lep2mcws} calculations
implementing full $\mathcal{O}(\alpha)$ electroweak radiative
corrections for 4-$f$ events in the so called Double Pole
Approximation (DPA)~\cite{dpa0,dpa1,dpa2}, i.e. reliable around the double
resonant $W$ pole, have been available as the result of an effort from
the theory community.  There are two Monte Carlo generators
implementing these calculations, {\tt YFSWW}~\cite{Yfsww} and {\tt
  RacoonWW}~\cite{RacoonWW}.

Initially the studies on the theoretical precision of these
calculations have been devoted to the inclusive $WW$ cross section,
showing a satisfactory 0.4\% agreement between the two codes. Studies
of differential distributions at generator level have been shown by
both the theoretical groups and by others (for instance~\cite{fabio}),
but a full attempt of assessing the theoretical precision on $W$
related observables has been presented only later for the $W$
mass~\cite{wmasssys} and for the TGC~\cite{tgcsys}.

In the TGC related study the possible sources of uncertainties in both
generators are considered and the calculations compared one to the
other. Moreover a detector effect parameterization (based on the ALEPH
simulation and analysis) is used to mimic the dominant effects beyond
the pure electroweak generator.

The $W$ mass study is a pure 4-$f + \gamma$ generator one based on a
pseudo-observable (the $\mu\nu$ mass with some photon recombination)
not directly comparable with the real observable measured by the
experiments. It is based on an internal precision study of {\tt YFSWW}
plus a comparison with {\tt RacoonWW}.

These studies provide a complete discussion of all the basic
ingredients of the systematic uncertainty related to electroweak
corrections, but the authors themselves recognize that for the $W$
mass a study at full analysis level is needed for a complete final
determination to be used by LEP experiments.

The purpose of the present work is to use the above mentioned studies
as a guideline to perform a complete estimation of this systematic
uncertainty for the $W$ mass analysis in the frame of the full DELPHI
event and detector simulation, reconstruction and analysis chain.  

In section~\ref{sec:delphi4f} the study of the intrinsic uncertainty
of the DELPHI 4-$f$ generator~\cite{delphi4f}, based on {\tt YFSWW} as
far as radiative corrections are concerned, is discussed. In
section~\ref{sec:racoonww} the comparison with {\tt RacoonWW} is
presented. Section~\ref{sec:results} shows the global results and
conclusions on the systematic uncertainty on the $W$ mass and width.

Although the target of the present study is the assessment of the
uncertainty on the $W$ mass, the techniques and the Monte Carlo
samples presented can be used for similar studies on other
observables, in particular the TGC.

\section{The uncertainty of the DELPHI 4-$f$ generator}
\label{sec:delphi4f}

\subsection{Description of the setups and samples}
\label{sec:delphisetup}

The 4-$f$ generator used for this study is the standard DELPHI one,
based on {\tt WPHACT}~\cite{wphact} with the YFS-exponentiated ISR
from {\tt KoralW}~\cite{KW1.42} and with additional radiative
corrections implemented for $WW$ like events through {\tt YFSWW},
using a reweighting technique as in the {\tt KandY} ``Monte Carlo
tandem''~\cite{KandY}: IBA based events are reweighted in order to
reproduce with good approximation the result of the DPA calculation.
For simplicity it will be referred to as {\tt WandY}. For single $W$
events and non $WW$-like final states an IBA approach is adopted,
using the {\tt QEDPS} parton shower generator~\cite{qedps} in order to
describe ISR, suitably adapted in the energy scale used for the
radiation.

The version used for this study, as well as for the final DELPHI $W$
mass analysis (internal DELPHI version 2.4) differs
from~\cite{delphi4f} in the treatment of the final state radiation
(FSR) from leptons, which is implemented with {\tt
  PHOTOS}~\cite{photos}: {\tt PHOTOS} version 2.5 is used,
implementing non leading logarithm (NLL) corrections which bring it
quite close to the full matrix element calculation~\cite{photosnll}.

The study has been performed at the centre of mass energy of $\sqrt{s}
= 188.6$ GeV, corresponding to the 1998 data sample. It has been
chosen since it represents the highest single-energy data statistics
available.

The wide range of sources of systematic uncertainties and possible
studies discussed in~\cite{wmasssys} implies the need for several
distinct Monte Carlo samples. Several sources can in fact be studied
by simple event reweighting, applying as event weight the ratio of the
modified matrix element squared and the standard one, where the
modifications are related to the uncertainty source to be studied. All
the possible weights have been implemented in the production of the
standard $WW$-like 4-$f$ samples.

Some studies cannot be performed by event reweighting and
require dedicated samples. In the standard {\tt WandY} the Leading
Pole Approximation (LPA) expansion around the double resonant pole
is made using the approach
that in {\tt YFSWW} is called the $\mbox{LPA}_A$ scheme~\cite{lpaa};
the other available approach, the so called $\mbox{LPA}_B$
scheme~\cite{lpab}, must be generated directly with {\tt YFSWW}.
Another case is the possible change of order in leptonic FSR: this
would require distinct samples with $\mathcal{O}(\alpha)$ and
$\mathcal{O}$$(\alpha^2)$ matrix elements.

Furthermore the need to compare {\tt WandY} to {\tt RacoonWW}, which
has some remarkable differences with respect to the normal DELPHI
code, has suggested to produce a dedicated {\tt WandY} sample suitably
modified to be as close as possible to {\tt RacoonWW} itself. Since
{\tt RacoonWW} cannot produce directly samples with several final
states at the same time, and the statistical precision needed for a
meaningful comparison ($\Delta m_W(\mbox{\tt Wandy - RacoonWW}) \simeq
\mathcal{O}$$(5 \, \mbox{MeV})$) requires about 1 million events per
channel to be produced, two final states have been chosen as
representatives of the fully hadronic and semileptonic channels for
these special event samples.

In order to minimize as much as possible 4-$f$ background
contamination to CC03 diagrams, CC11 final states have been selected;
the 4-$f$ background effect is better studied in the standard {\tt WandY}
sample, with massive kinematics and dedicated radiative corrections
not present in {\tt RacoonWW} and where inter-channel migration
effects, in which the 4-$f$ background can also play a role, can be
studied. For the fully hadronic channel the $udsc$ final state has
been chosen, and for the semileptonic channel $ud\mu\nu$ has been
preferred due to the presumably higher sensitivity to FSR corrections:
photons are likely to be seen, while in final states with electrons
most of them are merged in the calorimetric shower of the electron
itself, and in taus they are generally merged in the jet of particles
coming from the decay, which play a dominant role making all the
studies more complex.

In order to be directly comparable with {\tt RacoonWW}, these
dedicated samples have been produced with the following modifications
(compared to the standard settings):

\begin{itemize}
\item diagonal CKM matrix;
\item fixed $W$ and $Z$ widths;
\item $\mathcal{O}(\alpha)$ final state radiation from leptons with
  {\tt PHOTOS} version 2.5. It is closer to {\tt RacoonWW} than the
  original version in the lack of higher orders FSR;
\item no Coulomb correction, Khoze-Chapovsky ansatz Coulomb
  correction~\cite{KCansatz} implemented through reweighting.
\end{itemize}

Since in the normal production the standard Coulomb correction is
already included, the reweighting would allow to study only the
difference between this one and the approximated version of the full
non-factorizable $\mathcal{O}(\alpha)$ correction, the so called
Coulomb correction in the Khoze-Chapovsky ansatz. In order to study
the net $\mathcal{O}(\alpha)$ correction effect with respect to the
tree level (known to be significantly smaller than the previously
mentioned difference), no Coulomb correction is implemented in the
special samples generation.

The main concern of possible systematic differences in the results
from the dedicated samples and the standard ones is linked to the
propagators' width treatment. A test has been performed with a small
(100k events) dedicated $ud\mu\nu$ sample produced with the above
modification but the $W$ and $Z$ width, kept running. The $W$ mass
difference with respect to the main $ud\mu\nu$ sample was:
\begin{eqnarray}
  \Delta(\mbox{running} \, \Gamma_W - \mbox{fixed} \, \Gamma_W) &  = & -28 \pm 16
  \, \mbox{MeV}
\end{eqnarray}
well compatible with the known simple shift of $-27$ MeV of the mass
value when moving from the fixed to the running width
definition~\cite{lep2,runwidth}. This known shift has been verified at
generator level with a precision of about 2 MeV.

The {\tt WandY} code has been extensively compared to {\tt YFSWW}
(see~\cite{delphi4f}), and for CC03 events it has been shown to be
equivalent to {\tt KandY}. Anyway, as a further consistency cross
check, in order to allow the generalization of the results of this
study, a dedicated {\tt YFSWW} $udsc$ sample using $\mbox{LPA}_A$
scheme has been produced at pure ``4-$f + n~\gamma$'' level (including
FSR from quarks) to compare with a similar {\tt WandY} sample and with
{\tt RacoonWW} at a corresponding level. In appendix~\ref{sec:app1}
the input parameters set for {\tt YFSWW}, equivalent to what used in
{\tt WandY}, is given.

In the cross check only the CC03 part of {\tt WandY} has been used to
be consistent with {\tt YFSWW}. The total cross sections are found to
be in agreement at the ($0.03 \pm 0.06)~\%$ level. In the event analysis
photons forming an angle with the beam axis smaller than 2 degrees are
discarded, and those with a bigger angle are recombined with the
charged fermion with which they form the smallest invariant mass if
their energy is below 300 MeV or if this mass is below 5 GeV. Several
observables have been checked, among which the most interesting ones
for this study are invariant mass distributions. They have been fitted
using a fixed width like Breit-Wigner function:
\begin{eqnarray}
  BW(s) & = & \frac{P_3~s}{(s-(P_1+80.4)^2)^2+(P_1+80.4)^2P_2^2}
\label{eq:bw}
\end{eqnarray}
where the parameters $P_1$ and $P_2$ are the (shifted) $W$ mass and
width ($P_1$ actually represents the shift of the $W$ mass with
respect to 80.4 GeV/c$^2$). The absolute value obtained in the fit
depends on the fit function form and it is not particularly relevant.
What matters for this check is the level of agreement between
different codes when using the same analysis and fit procedure.

\begin{figure}[hbtp]
\begin{center}
  \scalebox{0.9}{\includegraphics{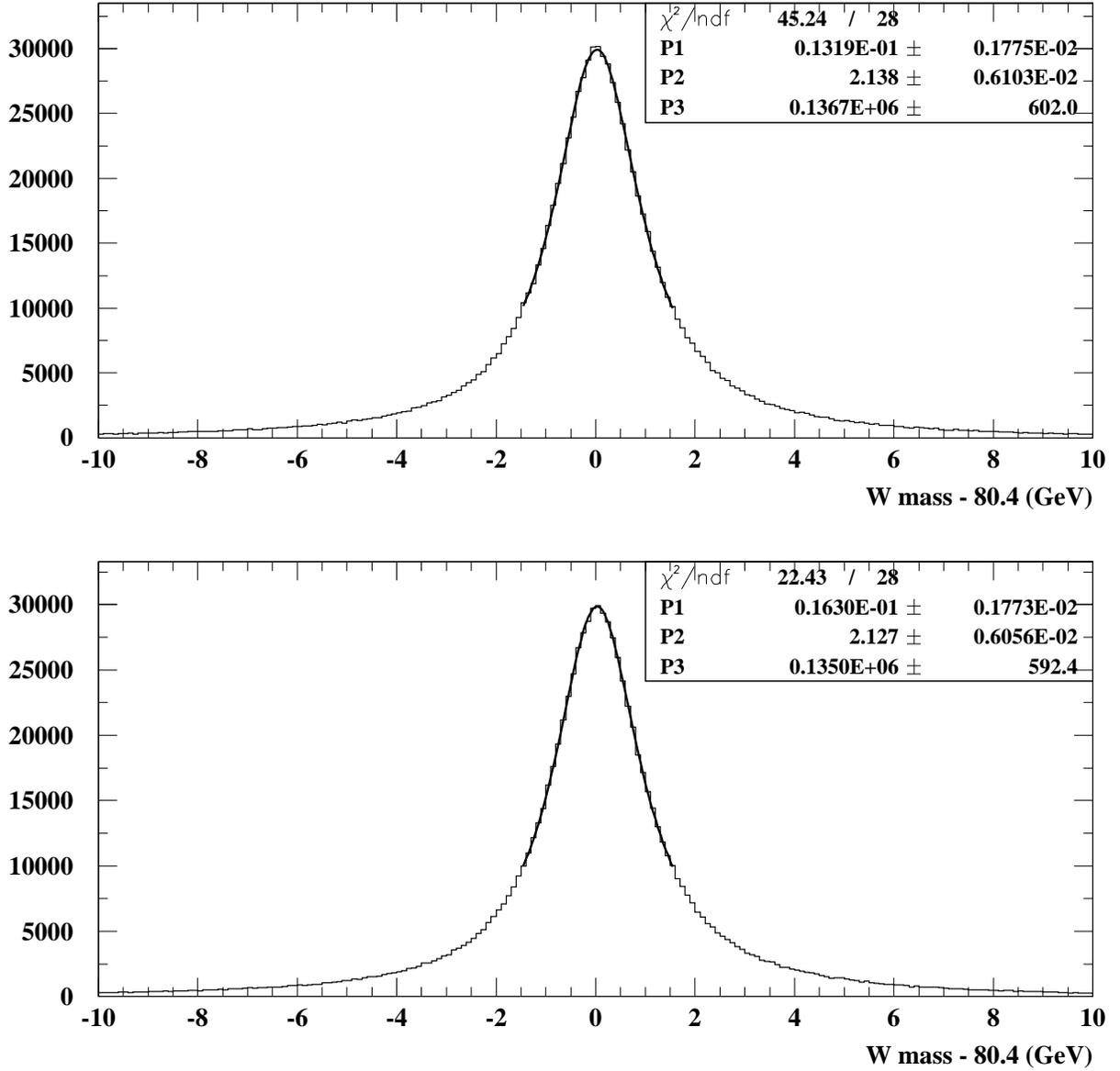}}
\end{center}
\caption{Average of $ud$ and $sc$ invariant masses (after photons cuts
  and recombination). Upper plot shows the result of a Breit-Wigner
  fit (eq.~\ref{eq:bw}) to the {\tt WandY} distribution, the lower one
  refers to {\tt YFSWW}.}
\label{fig1}
\end{figure}

\begin{figure}[hbtp]
\begin{center}
  \scalebox{0.9}{\includegraphics{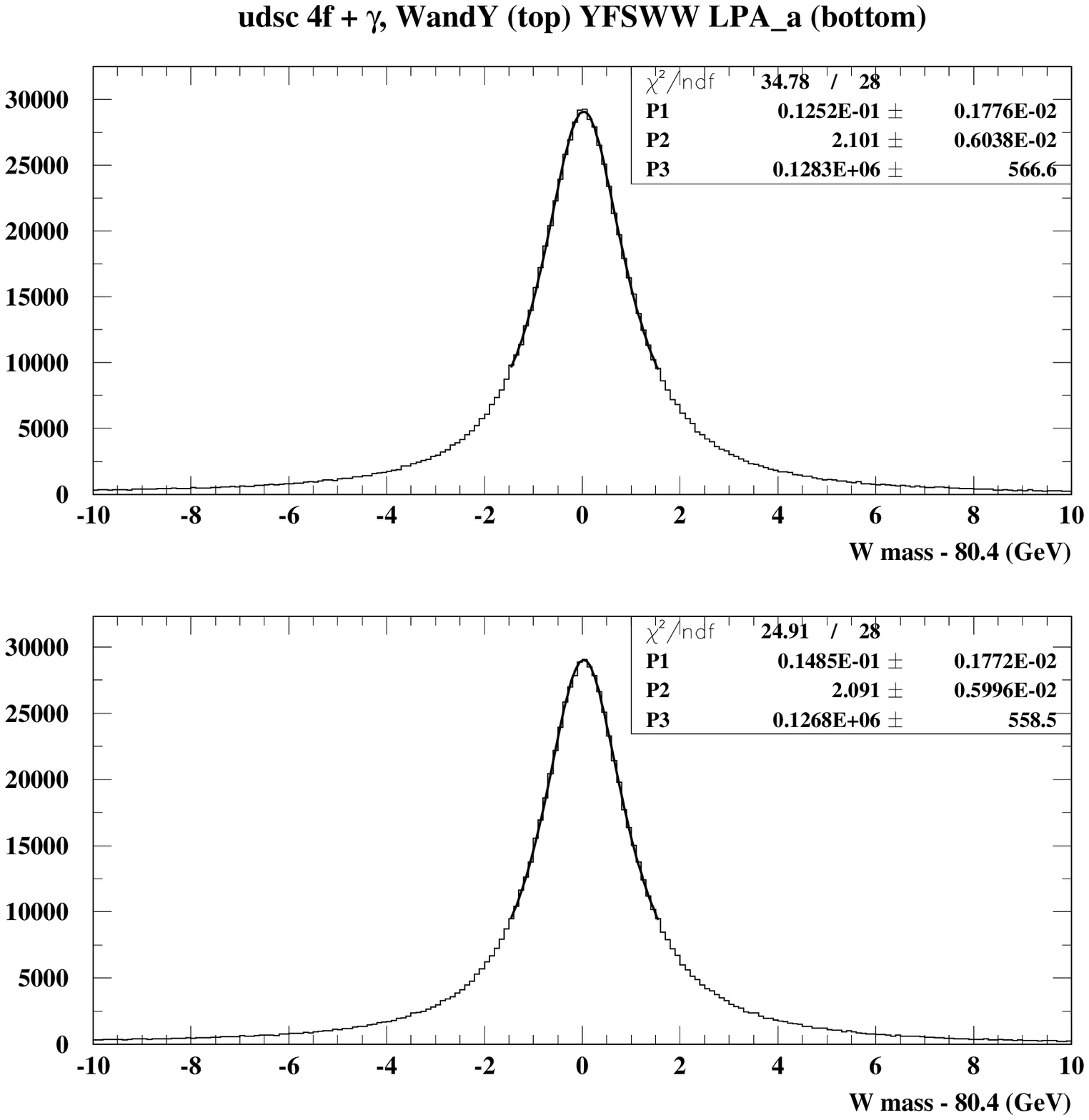}}
\end{center}
\caption{Average of the invariant masses obtained in the fermion
  pairing with the smallest masses difference (after photons cuts and
  recombination). Upper plot shows the result of a Breit-Wigner fit
  (eq.~\ref{eq:bw}) to the {\tt WandY} distribution, the lower one
  refers to {\tt YFSWW}.}
\label{fig2}
\end{figure}

\begin{figure}[hbtp]
\begin{center}
  \scalebox{0.9}{\includegraphics{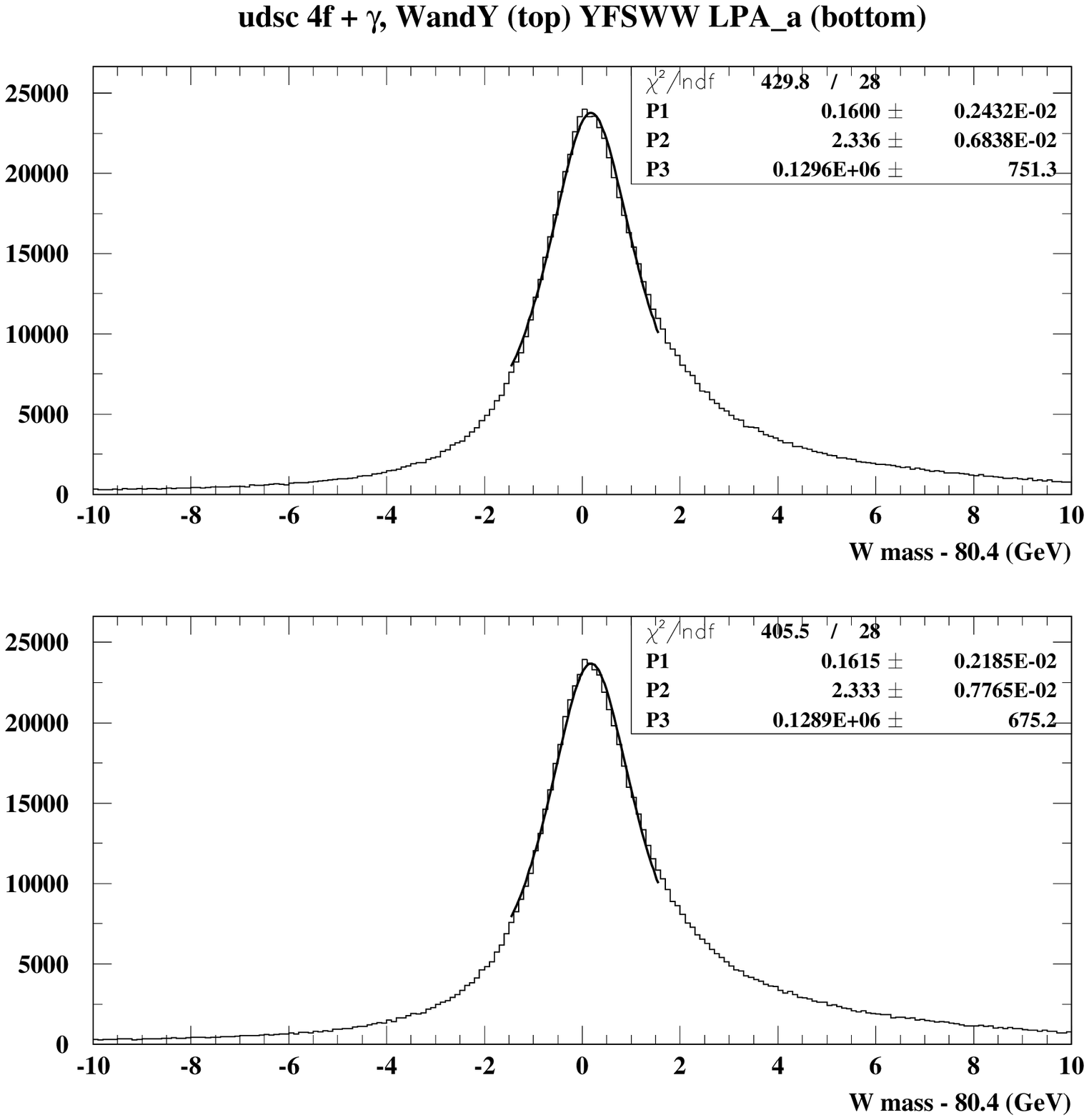}}
\end{center}
\caption{Average of the invariant masses obtained in the fermion
  pairing with the smallest masses difference after rescaling masses
  for the energies ratio $E_{beam}/E_{f\bar{f}}$ (after photons cuts
  and recombination). Upper plot shows the result of a Breit-Wigner
  fit (eq.~\ref{eq:bw}) to the {\tt WandY} distribution, the lower one
  refers to {\tt YFSWW}.}
\label{fig3}
\end{figure}

Fig.~\ref{fig1} shows the result of the fits on the average of the
$ud$ and $sc$ invariant masses. The agreement both in the mass and in
the width is satisfactory. An approach closer to the real analysis is
to look at the average of the masses from the pairing in which the
difference of the di-fermion masses is smallest (a criterion inspired by
the equal masses constraint used in constrained fits); the result is
shown in fig.~\ref{fig2}, and also here the agreement is good. An
observable that is very interesting, as will be seen in the comparison
with {\tt RacoonWW}, is the invariant mass rescaled by the ratio of
the beam energy and di-fermion energy: it is the simplest way to mimic
at pure generator level the energy-momentum conservation which is
usually imposed in constrained fits and which is responsible for
the sensitivity of the results to photon radiation, ISR in particular.
Differences in the radiation structure are likely to cause visible
effects in this kind of mass distributions, even if the previous ones
are in good agreement. In fig.~\ref{fig3} the average of the invariant
masses computed as in fig.~\ref{fig2} but rescaled by the ratio
$E_{beam}/E_{f\bar{f}}$ is shown: also in this case, despite the
sizeable effect of the rescaling on the fitted parameters compared to
the previous fits, the agreement is very satisfactory.

This check proves that the results based on {\tt WandY} can be
considered valid for similar analysis using {\tt YFSWW} (possibly
except for specific non CC03 diagrams related features).

\subsection{Technique of the uncertainty study}
\label{sec:systechnique}

The systematic uncertainty on the $W$ mass and width measurement due
to the electroweak radiative corrections is the effect of the
approximations and of the missing terms in the theoretical calculation
used for the analysis. Its exact knowledge would imply the full
computation of the missing corrections. The evaluation of the
systematic uncertainty means estimating the order of magnitude of the
effect of these not yet computed terms on the analysis.

This goal is practically achieved by splitting the calculations in
different parts (ISR, FSR, etc.), whose limited knowledge introduces a
source of uncertainty in the electroweak radiative corrections as
implemented in {\tt WandY}. The size of the uncertainty from each of
these sources can be estimated by repeating the full $W$ mass (and
width) analysis with changes in the part of the radiative corrections
related to this source, whose effect should reasonably be of the same
order of magnitude (or bigger) than the missing terms, and comparing
with the standard calculation. This study can be performed on both the
dedicated high statistics samples and on the standard ones.

The purely numerical precision from the fit algorithm is 0.1 MeV for
the mass value and 0.3 MeV for the mass error. On the width, due to
the very slow variation of the likelihood curve around the minimum,
the numerical accuracy on the fit result is about 1 MeV.

As already mentioned in the previous section, for several sources of
uncertainty it is possible to use a reweighting technique, which
allows to reuse the same event sample for several studies, minimizing
the simulation needed.  When using the reweighting technique, the
statistical error on the difference between the results of the fits on
the standard and the modified sample has to take into account the
correlation existing between the samples: the same events are used,
simply with a different weight in the fit. This correlation allows to
strongly reduce the error on the difference itself, with respect to
comparisons of statistically uncorrelated samples.

In order to take into account the correlation the total sample for one
channel has been divided into several subsamples, and the difference
has been computed for each subsample. The RMS of the subsamples
differences distribution, divided by the square root of the number of
subsamples, is an estimate of the uncertainty which naturally includes
the correlation between the original and reweighted samples. This way
of computing the errors has been cross checked for the mass (where
numerical fluctuations are generally negligible compared to the
statistical ones) with the ``Jackknife''~\cite{jackknife} one,
subtracting each time one subsample, and a very good agreement in the
error estimate has been found.

The study has been performed only on 4-$f$ $WW$-like events, omitting
all the remaining background processes. The rate and nature of the
total selected events which are discarded in this way strongly depends
on the channel~\cite{delphiwmass}:
\begin{list}{}{}
\item[$q\bar{q'}e\nu$]: $\simeq$ 5\%
\item[$q\bar{q'}\mu\nu$]: $<$ 1\%
\item[$q\bar{q'}\tau\nu$]: $\simeq$ 9\%
\item[$q\bar{q'}Q\bar{Q'}$]: $\simeq$ 24\%
\end{list}  
For semileptonic events they are both $q\bar{q'}ll$ and
$q\bar{q}\gamma$, the relative rate depending on the channel, while
for fully hadronic events practically only the latter class of events
weighs and is not considered. Other processes give anyway a negligible
contribution. The uncertainty from the radiative corrections on these
events is taken into account in the uncertainty on the background.

\subsection{Analysis of the sources of systematic uncertainties}
\label{sec:syssources}

Following the approach of ref.~\cite{wmasssys}, several distinct
categories of uncertainty sources common to all $WW$ channels can
be identified, corresponding to different parts of the electroweak
corrections:

\begin{itemize}{}{}
\item $WW$ production: initial state radiation (ISR);
\item $W$ decay: final state radiation (FSR);
\item Non-factorizable QED interference (NF) $\mathcal{O}(\alpha)$ corrections;
\item Ambiguities in LPA definition: non leading factorizable (NL)
  $\mathcal{O}(\alpha)$ corrections.
\end{itemize}

Moreover, due to the importance of the single $W$ diagrams in the
semileptonic electron channel and the relatively sizeable uncertainty
on the radiative corrections on them, a dedicated study has been
performed for semileptonic channels.

The uncertainty for each of the categories is studied by testing the
effect of activating/deactivating or modifying the relative
corrections, in order to have an estimate of the potential effect of
used approximations and non-calculated missing terms.

Table~\ref{tab:specmw} and~\ref{tab:specgw} show the results of the
studies for $m_W$ and $\Gamma_W$ respectively on the dedicated
samples, while table~\ref{tab:stdmw} and~\ref{tab:stdgw} show the
results on the standard samples.

\begin{table}[hbt]
\begin{center}
\begin{tabular}{|l|c|c|} 
\hline 
\multicolumn{3}{|c|}{$\Delta m_W$ (MeV)} \\
\hline \hline 
Numerical test & $ud\mu\nu$ & $udsc$ \\
\hline \hline
\multicolumn{3}{|c|}{Full DPA effect} \\
\hline 
Best - IBA &  $-10.6 \pm 0.7$ & $-10.1 \pm 1.0$ \\
\hline
\multicolumn{3}{|c|}{$WW$ production (ISR)} \\
\hline 
Best - $\mathcal{O}$$(\alpha^2)$ & $< -0.1$ & $< -0.1$ \\
Best - $\mathcal{O}(\alpha)$   & $-0.7 \pm 0.1$ & $-0.3 \pm 0.1$ \\ 
\hline
\multicolumn{3}{|c|}{$W$ decay (FSR)} \\
\hline 
Best - LL FSR                  & $< -0.1$ &      -      \\ 
\hline 
\multicolumn{3}{|c|}{Non-factorizable QED interference (NF $\mathcal{O}(\alpha)$)} \\
\hline
Best - no KC Coulomb           & $-0.7 \pm 0.1$ & $-1.9 \pm 1.0$ \\
\hline
\multicolumn{3}{|c|}{Ambiguities in LPA definition (NL $\mathcal{O}(\alpha)$)} \\
\hline
Best - EW scheme B             & $0.1 \pm 0.1$ & $< 0.1$ \\ 
Best - no NL ($\mbox{LPA}_A$)  & $-9.9 \pm 0.7$ & $-8.2 \pm 1.0$ \\ 
NL $\Delta (\mbox{no LPA}_A - \mbox{no LPA}_B)$ & $0.0 \pm 1.1$ &
    $1.3 \pm 1.0$ \\ 
\hline
\end{tabular}
\caption{Summary of the studies on the uncertainties on $m_W$
  performed on the dedicated $ud\mu\nu$ and $udsc$ samples. The quoted
  errors are statistical, and rounded to 0.1 MeV.}
\label{tab:specmw}
\end{center}
\end{table}

\begin{table}[hbt]
\begin{center}
\begin{tabular}{|l|c|c|} 
\hline
\multicolumn{3}{|c|}{$\Delta \Gamma_W$ (MeV)} \\
\hline \hline
Numerical test & $ud\mu\nu$ & $udsc$ \\
\hline \hline
\multicolumn{3}{|c|}{Full DPA effect} \\
\hline
Best - IBA &  $-9.4 \pm 1.4$ & $-17.0 \pm 1.0$ \\
\hline
\multicolumn{3}{|c|}{$WW$ production (ISR)} \\
\hline 
Best - $\mathcal{O}$$(\alpha^2)$ & $< -0.1$ & $< -0.1$ \\
Best - $\mathcal{O}(\alpha)$   & $-1.0 \pm 0.1$ & $-0.7 \pm 0.1$ \\ 
\hline
\multicolumn{3}{|c|}{$W$ decay (FSR)} \\
\hline 
Best - LL FSR                  & $-0.5 \pm 0.1$ &      -      \\ 
\hline 
\multicolumn{3}{|c|}{Non-factorizable QED interference (NF $\mathcal{O}(\alpha)$)} \\
\hline
Best - no KC Coulomb           & $1.6 \pm 0.1$ & $-0.4 \pm 0.1$ \\
\hline
\multicolumn{3}{|c|}{Ambiguities in LPA definition (NL $\mathcal{O}(\alpha)$)} \\
\hline
Best - EW scheme B             & $-0.1 \pm 0.1$ & $0.1 \pm 0.1$ \\ 
Best - no NL ($\mbox{LPA}_A$)  & $-11.1 \pm 1.4$ & $-16.6 \pm 1.0$ \\ 
NL $\Delta (\mbox{no LPA}_A - \mbox{no LPA}_B)$ & $3.9 \pm 2.8$ &
    $-1.6 \pm 4.0$ \\ 
\hline
\end{tabular}
\caption{Summary of the studies on the uncertainties on $\Gamma_W$
  performed on the dedicated $ud\mu\nu$ and $udsc$ samples. The quoted
  errors are statistical, and rounded to 0.1 MeV.}
\label{tab:specgw}
\end{center}
\end{table}

\subsubsection{$WW$ production: initial state radiation}

ISR is playing a key role in the $W$ mass analysis since it is one of
the main sources of the bias on the fit result with respect to the
true value, due to the energy-momentum conservation constraint used in
the kinematical constrained fits. The ISR is computed in the YFS
exponentiation approach, using a leading logarithm (LL)
$\mathcal{O}$$(\alpha^3)$ matrix element.

The difference between the best result, implementing the
$\mathcal{O}$$(\alpha^3)$ ISR matrix element and the
$\mathcal{O}$$(\alpha^2)$ one gives an order of magnitude of the effect
of the missing higher orders in the matrix element, i.e. to use a wrong
description of events with more than three hard photons or more than
one photon with high $p_t$. As can be seen from the tables, this
effect is below the fit sensitivity for all the channels.

The difference between the best result and the $\mathcal{O}(\alpha)$
includes the previous study, and can be used for estimating an upper
limit of the effect of the missing non leading logarithm (NLL) terms at
$\mathcal{O}$$(\alpha^2)$, which should be smaller than the LL
component removed. From the tables it is seen that the effect is
below 1 MeV both for the mass and the width in all the channels.

Taking into account also the study performed in~\cite{wmasssys}, the
ISR related uncertainty can be conservatively estimated at 1 MeV for
the mass and 2 MeV on the width.

\subsubsection{$W$ decay: final state radiation}

The FSR description and uncertainty is tightly linked to the final
state considered. QED FSR from quarks is embedded in the parton shower
describing the first phase of the hadronization process. It is
therefore essentially impossible to separate it from the rest of the
hadronization process, and the relative uncertainty is considered as
included in the jet and fragmentation related ones.

FSR from leptons is described by {\tt PHOTOS}. The difference between
the best result, based on the new NLL treatment, and the previous LL
one can give an estimate of the effect of the missing part of the
$\mathcal{O}(\alpha)$ FSR correction. It depends on the semileptonic
channel, but it is always within 1 MeV.

In~\cite{wmasssys} the effect of the missing higher orders beyond
$\mathcal{O}$$(\alpha^2)$ has been found to be negligible at generator level.
Since a full study of this uncertainty would require a high statistics
dedicated simulation, and simple perturbative QED considerations
suggest that the size of the effect should not exceed the size of the
previous one, conservatively the previous error can be doubled to take
into account also this component of the uncertainty.

\begin{table}[hbt]
\begin{center}
\begin{tabular}{|l|c|c|c|c|c|} 
\hline 
\multicolumn{6}{|c|}{$\Delta m_W$ (MeV)} \\
\hline 
\hline \rule{0cm}{0.5cm}
Numerical test & $q\bar{q'}e\nu$ & $q\bar{q'}\mu\nu$ & $q\bar{q'}\tau\nu$ & $q\bar{q'}l\nu$ & $q\bar{q'}Q\bar{Q'}$ \\
\hline \hline
\multicolumn{6}{|c|}{Full DPA effect} \\
\hline
Best - IBA &  $2.1 \pm 2.9$ & $6.3 \pm 2.0$ &  $1.6 \pm 3.4$ & $4.0 \pm 1.6$ & $5.6 \pm 1.0$ \\
\hline
\multicolumn{6}{|c|}{$WW$ production (ISR)} \\
\hline 
Best - $\mathcal{O}$$(\alpha^2)$ & $< -0.1$ & $< -0.1$ &  $< -0.1$ & $<
-0.1$ & $< -0.1$ \\
Best - $\mathcal{O}(\alpha)$   & $-0.8 \pm 0.1$ & $-0.6 \pm 0.1$ &  $-0.9 \pm 0.1$ & $-0.8 \pm 0.1$ & $-0.3 \pm 0.1$ \\ 
\hline
\multicolumn{6}{|c|}{$W$ decay (FSR)} \\
\hline 
Best - LL FSR                  & $< -0.1$ & $< -0.1$ & $-0.6 \pm
0.1$ & $-0.2 \pm 0.1$ &     -   \\ 
\hline 
\multicolumn{6}{|c|}{Non-factorizable QED interference (NF $\mathcal{O}(\alpha)$)} \\
\hline
Best - no KC Coulomb           & $16.5 \pm 0.2$ & $15.6 \pm 0.1$ &  $17.6 \pm 0.2$ & $16.3 \pm 0.1$ & $13.3 \pm 0.1$ \\
\hline
\multicolumn{6}{|c|}{Ambiguities in LPA definition (NL $\mathcal{O}(\alpha)$)} \\
\hline
Best - EW scheme B             & $0.2 \pm 0.1$ & $0.1 \pm 0.1$ &  $0.1 \pm 0.1$ & $0.1 \pm 0.1$ & $0.1 \pm 0.1$ \\ 
Best - no NL ($\mbox{LPA}_A$)  & $-14.4 \pm 2.9$ & $-9.6 \pm 2.0$ &  $-16.1 \pm 3.4$ & $-12.3 \pm 1.6$ & $-7.7 \pm 1.0$ \\ 
\hline
\end{tabular}
\caption{Summary of the studies on the uncertainties on $m_W$
  performed on the standard (all $WW$-like final states) sample. The
  quoted errors are statistical, and rounded to 0.1 MeV.}
\label{tab:stdmw}
\end{center}
\end{table}

\begin{table}[hbt]
\begin{center}
\begin{tabular}{|l|c|c|c|c|c|} 
\hline 
\multicolumn{6}{|c|}{$\Delta \Gamma_W$ (MeV)} \\
\hline 
\hline \rule{0cm}{0.5cm}
Numerical test & $q\bar{q'}e\nu$ & $q\bar{q'}\mu\nu$ & $q\bar{q'}\tau\nu$ & $q\bar{q'}l\nu$ & $q\bar{q'}Q\bar{Q'}$ \\
\hline \hline
\multicolumn{6}{|c|}{Full DPA effect} \\
\hline
Best - IBA &  $-16.3 \pm 7.7$ & $-17.7 \pm 5.3$ &  $-23.0 \pm 7.5$ & $-18.8 \pm 3.7$ & $-4.3 \pm 1.0$ \\
\hline
\multicolumn{6}{|c|}{$WW$ production (ISR)} \\
\hline 
Best - $\mathcal{O}$$(\alpha^2)$ & $< -0.1$ & $< -0.1$ &  $< -0.1$ & $< -0.1$ & $< -0.1$ \\
Best - $\mathcal{O}(\alpha)$   & $-1.0 \pm 0.1$ & $-1.0 \pm 0.1$ &
$-1.4 \pm 0.1$ & $-1.1 \pm 0.1$ & $-0.8 \pm 0.1$ \\ 
\hline
\multicolumn{6}{|c|}{$W$ decay (FSR)} \\
\hline 
Best - LL FSR                  & $-0.3 \pm 0.1$  & $-0.4 \pm 0.1$ & $-0.9 \pm
0.2$ & $-0.4 \pm 0.1$ &     -   \\ 
\hline 
\multicolumn{6}{|c|}{Non-factorizable QED interference (NF $\mathcal{O}(\alpha)$)} \\
\hline
Best - no KC Coulomb           & $-9.8 \pm 0.3$ & $-10.3 \pm 0.3$ &  $-10.2 \pm 0.4$ & $-9.7 \pm 0.2$ & $2.9 \pm 0.2$ \\
\hline
\multicolumn{6}{|c|}{Ambiguities in LPA definition (NL $\mathcal{O}(\alpha)$)} \\
\hline
Best - EW scheme B             & $-0.1 \pm 0.1$ & $-0.1 \pm 0.1$ &  $0.0 \pm 0.1$ & $-0.1 \pm 1.1$ & $0.1 \pm 0.1$ \\ 
Best - no NL ($\mbox{LPA}_A$) & $-6.8 \pm 7.7$ & $-7.9 \pm 5.3$ &  $-14.0 \pm 7.5$ & $-8.6 \pm 3.7$ & $-7.2 \pm 1.0$ \\ 
\hline
\end{tabular}
\caption{Summary of the studies on the uncertainties on $\Gamma_W$
  performed on the standard (all $WW$-like final states) sample. The
  quoted errors are statistical, and rounded to 0.1 MeV.}
\label{tab:stdgw}
\end{center}
\end{table}

\subsubsection{Non-factorizable QED interference: NF
  $\mathcal{O}(\alpha)$ corrections}

Non factorizable $\mathcal{O}(\alpha)$ corrections have to be
treated with care. It is known (see for
instance~\cite{fabio,wmasssys,KCansatz}) that the net effect of the
$\mathcal{O}(\alpha)$ QED interference between $W$s on the $W$ mass is
small if compared with Born level, and the apparent sizeable effect
seen when comparing new DPA calculations with the old IBA ones is an artifact
due to the use of the standard Coulomb correction.

This can be seen by comparing the results in tables~\ref{tab:specmw}
and~\ref{tab:specgw}, where the effective implementation of DPA NF
corrections through the Khoze-Chapovsky (KC) ansatz is compared to
the Born level (i.e. no correction at all), and the results in
tables~\ref{tab:stdmw} and~\ref{tab:stdgw}. Here the comparison is
done with the standard Coulomb correction, part of the traditional
IBA setup used before DPA.

The effect of using the KC ansatz with respect to Born can be considered as
an upper limit of the missing part of the full $\mathcal{O}(\alpha)$
calculation and of the higher order terms. Since the effect on the $W$
mass and width in comparing with the standard Coulomb correction on
all the final states is approximately the same for all the channels,
the values found on the special samples are used for all the final
states without further studies.

\subsubsection{Ambiguities in LPA definition: NL $\mathcal{O}(\alpha)$
  corrections}

The effect of the NL factorizable $\mathcal{O}(\alpha)$ corrections in
LPA is shown in all the tables. As it is seen, its almost complete
compensation with the change from standard Coulomb to KC Coulomb
correction is the reason for the small net effect of the full DPA
correction on the $W$ mass in comparison to the IBA. For the $W$ width
on the contrary the effects are in the same sense and add up.

Two sources of uncertainties are considered, following the study
in~\cite{wmasssys}. Missing higher orders effect can be, at least
partly, evaluated by changing the electroweak scheme used in the
$\mathcal{O}(\alpha)$ calculation. The
standard one in {\tt YFSWW} and {\tt WandY}, conventionally called A,
corresponds to the $G_{\mu}$ scheme, the other available one is called
B, and it corresponds to the choice of {\tt RacoonWW}. This
essentially means changing the definition of the 
QED fine structure constant used in the $\mathcal{O}(\alpha)$ matrix
element (see for instance the explanation in~\cite{Yfsww}). The effect is
very small, at the limit of the fit sensitivity, both for the mass and
the width.

It is worthwhile to notice here that in {\tt YFSWW} and {\tt WandY}
the $\mathcal{O}(\alpha)$ implementation beyond the standard IBA can
be technically splitted in two stages, the first one involving the
introduction of the WSR and ISR-WSR interference in the YFS form
factor and infrared $\tilde{S}$ factors, and the second one where the
electroweak virtual and soft $\mathcal{O}(\alpha)$ corrections and the
hard $\mathcal{O}(\alpha)$ matrix element are used to replace the pure
QED LL calculation. In this context it is interesting to notice that
the effect on the $W$ mass of the second phase is quite small when
compared to the total effect of the LPA correction, at most
$\mathcal{O}$$(5-10\%)$ of it. This allows to conclude
that the introduction of the ISR-WSR interference in the YFS form
factor and infrared $\tilde{S}$ factors plays a key role. For the $W$
width on the contrary the effect of the second part is found to be
much more important.

The second, more relevant, source of uncertainty connected to the LPA
is its possible definition, i.e. the ambiguity present in the way of
expanding the amplitude around the double resonant $W$ pole. The
standard {\tt YFSWW} and {\tt WandY} use the so called $\mbox{LPA}_A$
definition; a comparison with the $\mbox{LPA}_B$ one can give an
estimate of the effect from the instrinsic ambiguity in the LPA
definition.  Unfortunately $\mbox{LPA}_B$ cannot be reproduced through
reweighting, and it gives sizeable changes in comparison to
$\mbox{LPA}_A$ already at Born (or IBA) level. Therefore in order to
evaluate only the effect on the $\mathcal{O}(\alpha)$ correction a
separate $\mbox{LPA}_B$ sample has been generated with {\tt YFSWW},
and the effect has been estimated as the double difference:
\begin{eqnarray*}
  \Delta \mathcal{O}(\alpha) & \! \! \! (\mbox{LPA}_{A} - \mbox{LPA}_{B}) = & \Delta (\mbox{Best
  LPA}_A - \mbox{no NL LPA}_A) - \Delta (\mbox{Best
  LPA}_B - \mbox{no NL LPA}_B) 
\end{eqnarray*}
on the special samples. The size of the effect is within 1 MeV for the
mass, within 4 MeV for the width, dominated by the statistical uncertainty
(statistically independent samples are used). This result will be used
for all the final states and channels, since LPA is applied on the
CC03 part of the matrix element and therefore the estimate obtained
here should be approximately valid for all the final states.

\subsubsection{Radiative corrections on 4-$f$ background diagrams:
  single $W$}

At Born level the full 4-$f$ diagrams set for $WW$-like final states
is computed with a very high precision, at least for LEP2 energies and
in the phase space regions relevant for the $W$ mass and width
measurements. This was shown already by the studies in~\cite{lep2}.
Therefore the systematic uncertainties associated to it are linked
essentially to the electroweak corrections.

The DPA is known to be valid in a few $\Gamma_W$ interval around the
double resonant pole. The study of the previous section takes into
account the ambiguity in its definition and the effects caused by this
ambiguity far from the pole. Since the so called ``additive approach''
is used in {\tt WandY} for the DPA implementation through reweighting,
e.g. the DPA correction is applied only to the CC03 part of the matrix
element (and partly to the interference, see~\cite{delphi4f}), non
CC03 diagrams contributions are not directly affected by the DPA
uncertainty (except for possible effects in the interference term
which is relevant for the electron channel).

It is clear that this still leaves the problem of the approximated
radiative corrections treatment for the non CC03 part of the matrix
element (and the interference).

The ISR studies previously discussed can reasonably cover the most
relevant part of the electroweak radiative corrections uncertainties
present also for the $WW$-like 4-$f$ background diagrams, e.g. the
non CC03 part. There is a noticeable exception represented by the so
called single $W$ diagrams for the $q\bar{q'}e\nu$ final state
(see~\cite{lep2,lep2mcws} for their definition and a basic discussion
of the problem).

The bulk of single $W$ events is rejected in the $W$ mass and width
analysis, since the electron in these events is lost in the beam pipe.
But the CC03 - single $W$ interference is sizeable, and it has a strong
impact on the $W$ mass result in the electron channel.  This can be
easily seen from the variation of the $W$ mass result for the electron
channel when only the CC03 part of the matrix element is used in the
simulation (inter-final state cross talk is included):
\begin{eqnarray*}
  \Delta m_W \, (\mbox{electron}) \, \mbox{Best - CC03 only} & = & 106.6 \pm 1.9
  \, \mbox{MeV}
\end{eqnarray*}
and comparing with the variation when only the CC03/non CC03
interference is excluded from the simulation:
\begin{eqnarray*}
  \Delta m_W \, (\mbox{electron}) \, \mbox{Best - no interference} & = & 106.3 \pm 2.2
  \, \mbox{MeV}
\end{eqnarray*}

It can be noticed that the big effect of moving from a full 4-$f$
calculation to the CC03 only is almost entirely due to the
interference between the CC03 and the non CC03 part.

The situation is different in the $W$ width analysis, where in
$qqe\nu$ events reconstructed as electrons the effects of non CC03
diagrams and the CC03 - non CC03 interference are opposite in sign and
almost completely canceling.

The situation is made even more complex by the cross talk between
channel, e.g. events belonging in reality to one channel but
reconstructed as belonging to another one. This cross talk is
particularly relevant between electrons and taus, and this explains
why also the $\tau$ channel is sensitive to this uncertainty source.

The effect is particularly relevant for the width, where variations of
the non CC03 parts of the $qqe\nu$ matrix element give different
results with respect to the electron channel: the pure non CC03
diagrams give again an effect opposite in sign to the interference,
but much bigger, so in the width analysis the tau channel is more
sensitive to this systematic effect than the electron one:
\begin{eqnarray*}
  \Delta \Gamma_W \, (\mbox{tau}) \, \mbox{Best - CC03 only} & = & 190.7 \pm 12.3
  \, \mbox{MeV} \\
  \Delta \Gamma_W \, (\mbox{tau}) \, \mbox{Best - no interference} & = & -9.8 \pm 10.7
  \, \mbox{MeV} 
\end{eqnarray*}

Studying separately real $qq\tau\nu$ events from the $qqe\nu$ ones
reconstructed as taus clearly shows that this behaviour is due to the cross
talk.

Theoretical studies~\cite{lep2mcws} show that the standard IBA
calculations suffer from several problems for the
single $W$ process, ranging from gauge invariance issues to the scale
to be used for the ISR (the $t$-channel
scale should be preferred to the $s$-channel one), problems which can
globally lead to a $\mathcal{O}$$(4\%)$ uncertainty on the cross
section.

It should be noticed that {\tt WandY} implements several improvements
in this sector with respect to fixed width based IBA calculations
(see~\cite{delphi4f,wphact}). Nevertheless, in order to give an
estimate of the uncertainty related to the radiative corrections for
the single $W$ part, the non CC03 part of the matrix element, assumed
dominated by the single $W$ contribution, has been scaled by a factor
1.04 for $q\bar{q'}e\nu$ final states.

The effect on the mass and width measurement is shown in
table~\ref{tab:singlew}.

Another possible source of uncertainty related to 4-$f$ background is
represented by partly applying the DPA correction to the interference
term (see the discussion in~\cite{delphi4f}). The effect of this way
of computing the corrections is shown in table~\ref{tab:singlew}, and
can be considered as another estimate of the uncertainty related to
the 4-$f$ background presence.

\begin{table}[hbt]
\begin{center}
\begin{tabular}{|l|c|c|c|c|c|} 
\hline 
Numerical test & $q\bar{q'}e\nu$ & $q\bar{q'}\mu\nu$ &
$q\bar{q'}\tau\nu$ & $q\bar{q'}l\nu$ & $q\bar{q'}Q\bar{Q'}$ \\
\hline \hline 
\multicolumn{6}{|c|}{$\Delta m_W$ (MeV)} \\
\hline \hline
Best - non CC03 $\times$ 1.04 & $-4.2 \pm 0.1$ & $< -0.1$ & $0.6 \pm
0.1$ & $-1.2 \pm 0.1$ & - \\
Best - no DPA in int. & $-1.3 \pm 0.2$ & $0.2 \pm 0.1$ & $0.1 \pm 0.3$ &
$-0.3 \pm 0.1$ & $< 0.1$ \\
\hline \hline
\multicolumn{6}{|c|}{$\Delta \Gamma_W$ (MeV)} \\
\hline \hline
Best - non CC03 $\times$ 1.04 & $0.2 \pm 0.2$ & $< -0.1$ & $-6.4 \pm
0.4$ & $-1.2 \pm 0.1$ & - \\
Best - no DPA in int. & $1.8 \pm 0.5$ & $-0.4 \pm 0.1$ & $0.5 \pm 0.7$ &
$0.5 \pm 0.2$ & $< 0.1$ \\
\hline
\end{tabular}
\caption{Summary of the studies related to the uncertainties on $m_W$
  and $\Gamma_W$ due to 4-$f$ background radiative corrections
  performed on the standard (all $WW$-like final states) sample. The
  quoted errors are statistical, and rounded to 0.1 MeV.}
\label{tab:singlew}
\end{center}
\end{table}

\section{The DELPHI 4-$f$ generator - RacoonWW comparison}
\label{sec:racoonww}

The generator chosen by the LEP collaborations for implementing
electroweak radiative corrections in $WW$-like events is {\tt YFSWW},
used together with another full 4-$f$ generator (either {\tt KoralW} or
{\tt WPHACT}). {\tt RacoonWW} is the other, completely independent
Monte Carlo generator which implements radiative corrections in DPA on
top of a (massless) 4-$f$ generator.

Its use has been fundamental in assessing the DPA precision on the
$WW$ cross section, by comparing it with {\tt YFSWW}. It looks
therefore interesting to try to use it also for a completely
independent cross check of the {\tt YFSWW} based results on the $W$
mass and width (and possibly on other $W$ related measurements).  This
check has been already done in~\cite{wmasssys}, finding a good
agreement between the two codes, but as previously explained on an
observable which is not directly linked to the real analysis.

In appendix~\ref{sec:app2} the input options set used for {\tt
  RacoonWW} in this study is shown, and the output of one of the runs
is given to show the values of all the relevant parameters adopted for
the tuned comparison with {\tt WandY} and {\tt YFSWW}. The phase space
slicing approach has been adopted for the implementation of the
radiative corrections, in the version suggested for unweighted events
production ({\tt smc = 3}). The DELPHI version of {\tt PYTHIA} has
been used for the quark hadronization.

There is anyway a number of challenges in this test to be taken into
account. Real photon emission is handled in a completely different way
with respect to {\tt YFSWW}. In particular real emission in the
detector acceptance (i.e. with finite $p_t$) is computed only at
$\mathcal{O}(\alpha)$, although with a full 4-$f + \gamma$ matrix
element. Higher order ISR is present only through collinear structure
functions on events where there is no hard $\mathcal{O}(\alpha)$
emission, a very different situation compared to the YFS
exponentiation for ISR and WSR and the $\mathcal{O}$$(\alpha^3)$ LL ISR
matrix element. No FSR beyond the one already included in the
$\mathcal{O}(\alpha)$ is present, while in {\tt YFSWW} the FSR is
independent from the remaining part of the $\mathcal{O}(\alpha)$
calculation and introduced at $\mathcal{O}$$(\alpha^2)$ for leptons
through {\tt PHOTOS} and, merged with gluon emission, in the parton
shower for quarks. These differences have been investigated in the
literature (see for instance~\cite{lep2mcws,RacoonWW,fabio}) and are
known to give sizeable discrepancies in the photon related observables.

Therefore it is difficult to disentangle differences arising from a
different way of computing the same corrections from those due to the
use of different sets of corrections.

Since it is known that {\tt RacoonWW} in its DPA mode does not compare
well with {\tt YFSWW} on photonic spectra, the {\tt RacoonWW} authors
have developed a 4-$f + \gamma$ IBA mode which combines the
$\mathcal{O}(\alpha)$ matrix element and collinear structure
functions. The photonic energy and angular spectra produced in this
mode are in much better agreement with the {\tt YFSWW} ones at LEP2
energies, but it is not possible at present to combine it with the DPA
corrections for the virtual and soft emission part in a consistent
way.

Moreover the energy and angle cutoffs for the soft/hard photon
emission separation in {\tt RacoonWW} are in practice quite higher
than the {\tt YFSWW} ones, due to the quite different techniques
adopted in the two calculations. The phase space slicing approach for
matching virtual, soft and hard corrections has been used for this
test, and these cutoffs are an integral part of the approach itself.
The values used, shown in appendix~\ref{sec:app2}, correspond roughly
to a minimum real photon energy of about 95 MeV and a minimum real
photon-fermion angle of about 1.8 degrees, and are a compromise
between the reliability of the calculation and the attempt to avoid
merging with fermions photons which could be detected separately by
the detector. Moreover, in contrast to what has been suggested by the authors,
to avoid results which are dependent on the specific cutoff chosen, no
further photon recombination is applied in the sample production. This
choice is motivated by the fact that in a realistic simulation of a
detector any recombination has to be determined by the detector
granularity and analysis procedure itself, and due to the already big
values of the cutoffs adopted, any further recombination would risk to
suppress photons that would be detectable.

For final states with quarks, where the hadronization phase has to be
described beyond the electroweak radiative corrections, the use of a
full 4-$f + \gamma$ matrix element, in principle more correct than a
parton shower, creates in practice a problem: photons are
systematically emitted before gluons, which is unphysical and most
probably incompatible with the hadronization packages tunings used
({\tt PYTHIA}~\cite{pythia} is the standard choice for the analysis
and this study).

The suggestion of the authors of {\tt RacoonWW} to switch off the
photon radiation in the parton shower to compensate for the photon
emission in the matrix element has been adopted in this study, but it
does not seem a real solution to the problem, and of course it can
potentially spoil the validity of the hadronization tuning used. In
case of need this problem might be studied with the {\tt WandY} setup,
trying to emulate the {\tt RacoonWW} situation, i.e.  calling {\tt
  PHOTOS} also for quark pairs before the call to {\tt PYTHIA}, and
switching off photon emission inside {\tt PYTHIA} itself. This
presumably would overestimate the effect of FSR, since the photon
emission would be performed independently from the two fermion pairs.

A third potential problem in the comparison is represented by {\tt
  RacoonWW} generating massless fermions in the final state. Fermion
masses are added {\it a posteriori} using the routine provided by the
authors, which conserves obviously the total 4-momentum and the
di-fermion masses. It is clear that when a sizeable mass, compared to
the fermion energy, is added, as in the case of the $cs$ quark pair,
this could lead to distortions in the final state distributions.

All these features suggest that the comparison results must be
considered with care, if serious discrepancies are found (as it is the
case). On the other hand no special tuning has been prepared for the
hadronization package, in order to avoid mixing problems concerning
different sectors of the event description.

Table~\ref{tab:racoonww} shows the result of the comparison between
{\tt WandY} and {\tt RacoonWW 1.3}. A sizeable discrepancy can be seen
for the mass in the $ud\mu\nu$ channel, and, to a minor extent, for
the width in the $udsc$ channel.

\begin{table}[hbt]
\begin{center}
\begin{tabular}{|l|c|c|} 
\hline 
Numerical test & $ud\mu\nu$ & $udsc$ \\
\hline \hline 
\multicolumn{3}{|c|}{$\Delta m_W$ (MeV)} \\
\hline \hline
{\tt WandY} - {\tt RacoonWW 1.3} & $-38 \pm 5$ & $-4 \pm 5$ \\
\hline
\hline 
\multicolumn{3}{|c|}{$\Delta \Gamma_W$ (MeV)} \\
\hline \hline
{\tt WandY} - {\tt RacoonWW 1.3} & $4 \pm 10$ & $27 \pm 10$ \\
\hline
\end{tabular}
\caption{Summary of the {\tt WandY - RacoonWW} comparison on the
  uncertainties for $m_W$ and $\Gamma_W$ for the dedicated $ud\mu\nu$
  and $udsc$ final states. The quoted errors are statistical, and
  rounded to 1 MeV.}
\label{tab:racoonww}
\end{center}
\end{table}

Extensive studies have been performed in order to investigate the
discrepancies, in particular the one on the $W$ mass.

The different hadronization due to the treatment of FSR from quarks in
{\tt RacoonWW} has of course an influence on the jet characteristics, and
can affect the results, in particular the ones for the width.
Optimizing the interface of the hadronization with the electroweak
full matrix element to circumvent possible problems arising from the
simple minded approach followed goes beyond the scope of this study.

A generator level analysis analogous to the one whose results are
shown in fig.~\ref{fig1},~\ref{fig2} and~\ref{fig3}, has been used for
a 4-$f + \gamma$ level comparison of {\tt WandY} with {\tt RacoonWW
  1.3} for the $ud\mu\nu$ channel (all the 4-$f$ diagrams are included
here, not only the CC03 part). This study has been used to investigate
the discrepancy on the $W$ mass trying to disentangle the genuine
electroweak part from possible problems connected to the
implementation of the hadronization phase.

This study has clearly shown the crucial role played by the photon
clustering, in particular around the muon. The different treatment
of the soft, but mainly of the collinear photons in the two codes
implies a strong difference in the radiation around the fermions. In
{\tt RacoonWW} no visible photon is generated in a cone of 1.8 degrees
around a fermion, no matter which energy it has, and the radiation is
reassociated to the lepton. This is not true in {\tt WandY}, were the
energy and angle cutoffs (for FSR from leptons the {\tt PHOTOS} ones)
are quite smaller, closer to a real situation.

For quarks this is not a big problem since experimentally FSR photons
cannot be disentangled from jets, and they are naturally clustered
to the jets themselves. But the treatment of photons around leptons is
a different problem. While in the reconstruction of high energy
electrons a clustering of photons is done in order to take into
account the bremsstrahlung due to the interaction with the detector,
muons can be quite cleanly separated from photons, unless they are
strictly collinear. In the latter case the photon energy is anyway
lost, since the muon momentum is used, not the energy deposited in the
calorimeters possibly associated to it. $ud\mu\nu$ is therefore a good
final state to study in detail differences in the visible photon
radiation, mainly FSR.

In the real analysis visible photons, which have passed the quality
selection criteria, are clustered to the muon if in a cone of 3
degrees around it, otherwise are associated to the jets.  This
procedure can partly reabsorb the difference in the collinear
radiation mentioned above, even if not completely, because of limited
photon reconstruction efficiency, resolution, selection cutoffs, etc.
The effect of this photon clustering is of improving the agreement
between the two calculations on the fitted mass,
without it the difference in table~\ref{tab:racoonww} would be about
-50 MeV.

The $W$ mass difference obtained on the beam energy rescaled average
mass (like in fig.~\ref{fig3}) is -6 MeV if photons are clustered to
the charged fermion with which they have the smallest $p_t$. If on the
contrary the clustering to the muon is done only for photons in a
3 degrees cone around it, associating all the others to the quarks,
the difference becomes -23 MeV.

Increasing the opening angle of the cone for the clustering improves
the agreement, but of course in the real analysis such a procedure
would rapidly cluster photons coming from the hadronization of the quarks
(mainly $\pi^0$ decay products). Although the opening angle might be
tuned to minimize the rate of photons from jets clustered and
optimize the {\tt WandY} - {\tt RacoonWW} agreement, such a procedure
would introduce further systematic uncertainties due to the imperfect
knowledge of the photon distributions in jets. 

The residual discrepancy is presumably linked to the known differences
between the two calculations in the description of the radiation
beyond the treatment of the strictly collinear region in this study. The
good agreement for the mass found in the hadronic channel seems due to
the smaller sensitivity of the analysis to the detailed description of
the photonic radiation, since the photon clustering is implicit in
the analysis procedure itself. This looks anyway an encouraging result
for the general confidence in the study.

In this situation using the difference between the prediction of the
two calculations to estimate the systematic uncertainty on the $W$
mass and width does not seem appropriate. 

\section{Results and conclusions}
\label{sec:results}

The results of all the studies presented have to be combined in a
single uncertainty for each channel. Tables~\ref{tab:deltamw}
and~\ref{tab:deltagw} present an estimate of the different sources
of uncertainties as it can be deduced from the studies presented in
the section~\ref{sec:delphisetup}. Where the numerical or statistical
uncertainty on the estimate is comparable with the estimate itself,
they are added linearly to take them conservatively into account.

\begin{table}[hbt]
\begin{center}
\begin{tabular}{|l|c|c|c|c|} 
\hline 
\multicolumn{5}{|c|}{$\Delta m_W$ (MeV)} \\
\hline \hline
Uncertainty source & $q\bar{q'}e\nu$ & $q\bar{q'}\mu\nu$ & $q\bar{q'}\tau\nu$ & $q\bar{q'}Q\bar{Q'}$ \\
\hline
ISR                      & 1 & 1 & 1 & 1 \\
FSR                      & 0.5 & 0.5 & 1 & - \\
NF $\mathcal{O}(\alpha)$ & 1 & 1 & 1 & 2 \\
NL $\mathcal{O}(\alpha)$ & 1 & 1 & 1 & 1  \\
4-$f$ background         & 5.5 & 0.5 & 1 & 0.5 \\   
\hline
Total & 9 & 4 & 5 & 4.5 \\
\hline
\end{tabular}
\caption{Summary of the systematic uncertainties on the $W$ mass. The
  total is computed adding linearly the values of all the contributions.}
\label{tab:deltamw}
\end{center}
\end{table}

\begin{table}[hbt]
\begin{center}
\begin{tabular}{|l|c|c|c|c|} 
\hline 
\multicolumn{5}{|c|}{$\Delta \Gamma_W$ (MeV)} \\
\hline \hline
Uncertainty source & $q\bar{q'}e\nu$ & $q\bar{q'}\mu\nu$ & $q\bar{q'}\tau\nu$ & $q\bar{q'}Q\bar{Q'}$ \\
\hline
ISR                      & 2 & 2 & 2 & 2 \\
FSR                      & 1 & 1 & 2 & - \\
NF $\mathcal{O}(\alpha)$ & 2    & 2 & 2 & 2 \\
NL $\mathcal{O}(\alpha)$ & 4    & 4 & 4 & 4  \\
4-$f$ background         & 2  & 1 & 6 & 1 \\   
\hline
Total & 11 & 10 & 16 & 9 \\
\hline
\end{tabular}
\caption{Summary of the systematic uncertainties on the $W$ width. The
  total is computed adding linearly the values of all the contributions.}
\label{tab:deltagw}
\end{center}
\end{table}

The total uncertainty per channel is computed summing linearly the
values of the contributions. This choice is conservatively
motivated by the fact that several contributions are more maximal
upper limits than statistical errors. All the numbers have been
rounded to 0.5 MeV.

As can be seen, the uncertainty on the $W$ mass is within the 10
MeV level.

\section{Acknowledgments}

I am grateful to the authors of {\tt YFSWW}, S.~Jadach, W.~Placzek,
M.~Skrzypek, B.~F.~L.~Ward and Z.~Was, and to those of {\tt RacoonWW},
A.~Denner, S.~Dittmaier, M.~Roth and D.~Wackeroth, for the useful
discussions, the help in understanding both the theoretical problems
connected to this study and the way of correctly performing it. I
would also like to thank the authors of {\tt RacoonWW} for the help in
cross checking my results. I want to thank the DELPHI colleagues
involved in the $W$ mass measurement for the useful feedback and for
reading this manuscript, and in particular J.~D'Hondt for extracting
the results of this study for the fully hadronic channel. 
Finally I want to thank A.~Ballestrero and R.~Chierici for the fruitful
long lasting discussions and exchange of ideas on these problems.

\newpage
\appendix
\section{Appendix: {\tt YFSWW} input parameters}
\label{sec:app1}

The {\tt YFSWW} samples used for the study, whose settings are the
same as those used in {\tt WandY} special samples, have been generated
with version 3-1.17. The input for the $LPA_A$ sample ($udsc$ final
state) is:

\begin{verbatim}
*//////////////////////////////////////////////////////////////////////////////
*//                                                                          //
*//                 Input data for YFSWW3: ISR + EW + FSR                    //
*//                     For Simple DEMO Program                              //
*//                                                                          //
*//////////////////////////////////////////////////////////////////////////////
BeginX
*<-i><----data-----><-------------------comments------------------------------>
    1        188.6d0 CMSEne =xpar( 1) ! CMS total energy [GeV]
    2     1.16639d-5 Gmu    =xpar( 2) ! Fermi Constant  
    4       91.187d0 aMaZ   =xpar( 4) ! Z mass   
    5     2.506693d0 GammZ  =xpar( 5) ! Z width      
    6       80.400d0 aMaW   =xpar( 6) ! W mass 
    7     -2.08699d0 GammW  =xpar( 7) ! W with, For gammW<0 it is RECALCULATED
   11          115d0 amh    =xpar(11) ! Higgs mass 
   13       0.1255d0 alpha_s=xpar(13) ! QCD coupling const.
  111            1d0 vckm(1:1)
  112            0d0 vckm(1:2)
  113            0d0 vckm(1:3)
  114            0d0 vckm(2:1)
  115            1d0 vckm(2:2)
  116            0d0 vckm(2:3)
  117            0d0 vckm(3:1)
  118            0d0 vckm(3:2)
  119            1d0 vckm(3:3)
*<-i><----data-----><-------------------comments------------------------------>
* YFSWW3 SPECIFIC PARAMETERS !!!
*=============================================================================
 2001            5d0 KeyCor =xpar(2001)   Radiative Correction switch
*                    KeyCor   =0: Born
*                             =1: Above + ISR
*                             =2: Above + Coulomb Correction
*                             =3: Above + YFS Full Form-Factor Correction
*                             =4: Above + Radiation from WW
*                             =5: Above + Exact O(alpha) EWRC (BEST!)
*                             =6: As Above but Apporoximate EWRC (faster) 
 2002            0d0 KeyLPA   =0: LPA_a
*=============================================================================
 1011            1d0 KeyISR  =0,1  initial state radiation off/on (default=1)
*
 1013            1d0 KeyNLL  =0 sets next-to leading alpha/pi terms to zero
*                            =1 alpha/pi in yfs formfactor is kept (default)
 1014            2d0 KeyCul =xpar(1014)
*                    =0 No Coulomb correction 
*                    =1 "Normal" Coulomb correction 
*                    =2 "Screened-Coulomb" Ansatz for Non-Factorizable Corr. 
 1021            2d0 KeyBra =xpar(1021)
*                    = 0 Born branching ratios, no mixing
*                    = 1 branching ratios from input
*                    = 2 branching ratios with mixing and naive QCD 
*                       calculated in IBA from the CKM matrix (PDG 2000); 
*                       see routine filexp for more details (file filexp.f)
 1023            1d0 KeyZet =xpar(1023)
*                    = 0, Z width in z propagator: s/m_z *gamm_z
*                    = 1, Z width in z propagator:   m_z *gamm_z
*                    = 2, Z zero width in z propagator.
 1026            1d0 KeyWu  =xpar(1026)
*                    = 0 w width in w propagator: s/m_w *gamm_w
*                    = 1 w width in w propagator:   m_w *gamm_w
*                    = 2 no (0) w width in w propagator.
 1031            0d0 KeyWgt =xpar(1031)
*                    =0, unweighted events (wt=1), for apparatus Monte Carlo
*                    =1, weighted events, option faster and safer
 1041            1d0 KeyMix =xpar(1041)
*                    KeyMix EW "Input Parameter Scheme" choices. 
*                    =0 "LEP2 Workshop '95" scheme (for Born and ISR only!)
*                    =1 G_mu scheme (RECOMMENDED)
* W decays: 1=ud, 2=cd, 3=us, 4=cs, 5=ub, 6=cb, 7=e, 8=mu, 9=tau, 0=all chan.
 1055            4d0 KeyDWm =xpar(1055)    W- decay: 7=(ev), 0=all ch.   
 1056            1d0 KeyDWp =xpar(1056)    W+ decay: 7=(ev), 0=all ch.  
 1057           16d0 Nout   =xpar(1057)    Output unit no, for Nout<0, Nout=16
*============================================================================= 
*                  TAUOLA, PHOTOS, JETSET
*   >>> If you want to switch them OFF, uncomment the lines below <<< 
*<-i><----data-----><-------------------comments------------------------------>
 1071            0d0 Jak1   =xpar(1071)   Decay mode tau+
 1072            0d0 Jak2   =xpar(1072)   Decay mode tau-
 1073            0d0 Itdkrc =xpar(1073)   Bremsstrahlung switch in Tauola 
 1074            2d0 IfPhot =xpar(1074)   PHOTOS switch
 1075            0d0 IfHadM =xpar(1075)   Hadronization W-
 1076            0d0 IfHadP =xpar(1076)   Hadronization W+
  516        0.01d0       mass [GeV] (3-9 MeV in PDG)
  526        0.005d0      mass [GeV] (1.5-5 MeV in PDG)
  536        0.2d0        mass [GeV] (60-170 MeV in PDG)
  546        1.30d0       mass [GeV] (1.1-1.4 GeV in PDG)
  556        4.8d0        mass [GeV] (4.1-4.4 GeV in PDG)
  566        175.0d0      mass [GeV] (174.3 GeV in PDG 1999)
EndX
*//////////////////////////////////////////////////////////////////////////////
\end{verbatim}

The $LPA_b$ sample input, used for fully simulated events, differed in
the following parameters:

\begin{verbatim}
 2002            1d0 KeyLPA   =0: LPA_a
 1073            1d0 Itdkrc =xpar(1073)   Bremsstrahlung switch in Tauola 
 1074            1d0 IfPhot =xpar(1074)   PHOTOS switch
 1075            1d0 IfHadM =xpar(1075)   Hadronization W-
 1076            1d0 IfHadP =xpar(1076)   Hadronization W+
\end{verbatim}

The {\tt YFSWW} version used for the full simulation implemented the
{\tt PYTHIA} version and tuning and the {\tt TAUOLA} version used
in~\cite{delphi4f}.

\newpage
\section{Appendix: {\tt RacoonWW} input options and parameters}
\label{sec:app2}

The {\tt RacoonWW} samples used for the study, with input options and
parameters tuned to give the best agreement with the {\tt WandY} and
{\tt YFSWW} samples used, have been generated both with version 1.2
and 1.3. The input for the 1.3 sample ($udsc$ final state) is:

\begin{verbatim}
udsc.out    ! name of output file
188.6d0     ! energy: CMF energy (in GeV)
100000      ! neventsw: number of weighted events
3           ! smc: choice of MC branch: 1(or 3):slicing 2:subtraction
1           ! sborn4: include Born ee->4f: 0:no 1-3:yes
1           ! sborn5: include Born ee->4f+photon: 0:no 1:yes
0           ! sborng5: include Born ee->4f+gluon: 0:no 1:yes
1           ! sisr: include higher-order ISR: 0:no 1:yes
1           ! src: include radiative corrections: 0:no 1:DPA 2:IBA-4f 3:IBA-4fa
0           ! scoul5: Coulomb singularity for ee->4f+photon: 0:no 1,2:yes
3           ! qnf: Coulomb singularity for ee->4f: 1,2, or 3
0           ! qreal: neglect imaginary part of virt. corr.: 0:no 1:yes
2           ! qalp: choice of input-parameter scheme: 0,1, or 2
4           ! qgw: calculate the W-boson width: 0:no 1-4:yes
1           ! qprop: choice of width scheme: 0,1,2,3 or 4 
0           ! ssigepem4: choice of diag. for Born ee->4f: 0:all 1-5:subsets
5           ! ssigepem5: choice of diag. for Born ee->4f+ga: 0:all 1-5:subsets
0           ! ssigepemg5: choice of diag. for Born ee->4f+gl: 0:all 1,5:subsets
2           ! qqcd: include QCD radiative corr.: 0:no 1:CC03 2:naive 3:CC11
0           ! sqcdepem: include gluon-exch. diag. in Born: 0:no 1:yes 2:only
u           ! fermion 3
d           ! anti-fermion 4
s           ! fermion 5
c           ! anti-fermion 6
0d0         ! pp: degree of positron beam polarization [-1d0:1d0]
0d0         ! pm: degree of electron beam polarization [-1d0:1d0]
0           ! srecomb: recombination cuts: 0:no 1:TH 2:EXP
1.7d0       ! precomb(1): angular rec. cut between photon and beam
0.1d0       ! precomb(2): rec. cut on photon energy
1.32d0      ! precomb(3): inv.-mass rec.(TH) or angular rec. cut for lept.(EXP)
0d0         ! precomb(4): angular rec. cut for quarks(EXP)
0           ! srecombg: gluon recombination cuts: 0:no 1:TH 2:EXP
0d0         ! precombg(1): rec. cut on gluon energy
0d0         ! precombg(2): inv.-mass (TH) or angular (EXP) recombination cut 
0           ! satgc: anomalous triple gauge couplings (TGC): 0:no 1:yes
0d0         ! TGC Delta g_1^A
0d0         ! TGC Delta g_1^Z
0d0         ! TGC Delta kappa^A
0d0         ! TGC Delta kappa^Z
0d0         ! TGC lambda^A
0d0         ! TGC lambda^Z
0d0         ! TGC g_4^A
0d0         ! TGC g_4^Z
0d0         ! TGC g_5^A
0d0         ! TGC g_5^Z
0d0         ! TGC tilde kappa^A
0d0         ! TGC tilde kappa^Z
0d0         ! TGC tilde lambda^A
0d0         ! TGC tilde lambda^Z
0d0         ! TGC f_4^A
0d0         ! TGC f_4^Z
0d0         ! TGC f_5^A
0d0         ! TGC f_5^Z
0d0         ! TGC h_1^A
0d0         ! TGC h_1^Z
0d0         ! TGC h_3^A
0d0         ! TGC h_3^Z
0           ! qaqgc: anomalous quartic gauge couplings (QGC): 0:no 1:yes
0d0         ! QGC a_0/Lambda^2
0d0         ! QGC a_c/Lambda^2  
0d0         ! QGC a_n/Lambda^2  
0d0         ! QGC tilde a_0/Lambda^2
0d0         ! QGC tilde a_n/Lambda^2
10          ! scuts: separation cuts: 0:no 1,2:default(ADLO,LC) 10,11:input
0d0         ! photon(gluon) energy cut
1d0         ! charged-lepton energy cut
2d0         ! quark energy cut
2d0         ! quark-quark invariant mass cut
0d0         ! angular cut between photon and beam
0d0         ! angular cut between photon and charged lepton
0d0         ! angular cut between photon(gluon) and quark
0d0         ! angular cut between charged leptons
0d0         ! angular cut between quarks
0d0         ! angular cut between charged lepton and quark
0d0         ! angular cut between charged lepton and beam
0d0         ! angular cut between quark and beam
\end{verbatim}

and the corresponding output is:

\begin{verbatim}
  
      smc= 3: Phase-space-slicing branch of RacoonWW
              ======================================
 
 technical cutoff parameters (photon): 
  delta_s =    1.0000000000000000E-003
  delta_c =    5.0000000000000000E-004

 Input parameters:
 -----------------
    CMF energy = 188.60000 GeV,    Number of events =     100000,

  alpha(0) = 1/ 137.0359895, alpha(MZ) = 1/128.88700,   alpha_s = 0.12550,
        GF = .1166390E-04,
        MW =  80.40000,             MZ =  91.18700,          MH = 115.00000,
        GW =   2.09372,             GZ =   2.50669,
        me = .51099906E-03,        mmu =   0.105658300,    mtau =   1.77700,
        mu =   0.00500,             mc =   1.30000,          mt = 175.00000,
        md =   0.01000,             ms =   0.20000,          mb =   4.80000.

 Effective branching ratios: 
 leptonic BR =  0.32476, hadronic BR =  0.67524, total BR =  1.00000
  
 Process:  anti-e e -> u anti-d s anti-c (+ photon)                   
  
       pp= 0.0: degree of positron beam polarization.
       pm= 0.0: degree of electron beam polarization.
     qalp= 2: GF-parametrization scheme.
      qgw= 4: one-loop W-boson width calculated (with QCD corr.).
    qprop= 1: constant width.
  
   sborn4= 1: tree-level process ee -> 4f.
ssigepem4= 0: all electroweak diagrams included.
     qqcd= 2: naive QCD corrections included. 
  
      src= 1: virtual corrections in DPA and real corrections included.
ssigepem5= 5: real photon corr. : only CC11 class of diagrams included.
     qqcd= 2: naive QCD corrections included 
    qreal= 0: imaginary part of virtual corrections included.
      qnf= 3: off-shell Coulomb singularity with off-shell Born included.
     sisr= 1: initial-state radiation up to order alpha^3 included.
  
    scuts=10: with separation cuts:
 energy(3)  >   2.00000 GeV
 energy(4)  >   2.00000 GeV
 energy(5)  >   2.00000 GeV
 energy(6)  >   2.00000 GeV
 mass(3,4)  >   2.00000 GeV
 mass(3,5)  >   2.00000 GeV
 mass(3,6)  >   2.00000 GeV
 mass(4,5)  >   2.00000 GeV
 mass(4,6)  >   2.00000 GeV
 mass(5,6)  >   2.00000 GeV
  
    events  :  intermediary results   :  preliminary results 
   1000000  : 1846.54692 +- 11.43978  : 1846.54692 +- 11.43978
   2000000  : 1833.80408 +- 11.11287  : 1840.17550 +-  7.97440
   3000000  : 1852.08809 +- 11.09372  : 1844.14636 +-  6.47590
   4000000  : 1839.58697 +- 11.05121  : 1843.00651 +-  5.58773
   5000000  : 1834.96514 +- 10.89130  : 1841.39824 +-  4.97266
   6000000  : 1843.06512 +- 11.03453  : 1841.67605 +-  4.53366
   7000000  : 1840.81013 +- 10.93956  : 1841.55235 +-  4.18847
   8000000  : 1832.80339 +- 10.87736  : 1840.45873 +-  3.90900
 Warning: weight=-1            1       19685
   9000000  : 1836.94865 +- 10.94773  : 1840.06872 +-  3.68143
Warning: weight>weighttotmax           1       21733
         weight/weighttotmax=.21913D+01
         Redefining weighttotmax=weight
  10000000  : 1844.61537 +- 11.08916  : 1840.52339 +-  3.49394
  11000000  : 1851.84340 +- 11.08708  : 1841.55248 +-  3.33239
  12000000  : 1841.10316 +- 10.94892  : 1841.51504 +-  3.18804
  13000000  : 1837.82755 +- 10.94003  : 1841.23138 +-  3.06077
  14000000  : 1846.96380 +- 10.97894  : 1841.64084 +-  2.94835
  15000000  : 1830.80012 +- 10.84000  : 1840.91813 +-  2.84510
  16000000  : 1854.14538 +- 11.11172  : 1841.74483 +-  2.75621
  17000000  : 1841.29479 +- 10.91541  : 1841.71836 +-  2.67237
  18000000  : 1841.08998 +- 10.99218  : 1841.68345 +-  2.59673
  19000000  : 1844.23691 +- 10.97313  : 1841.81784 +-  2.52694
  20000000  : 1859.00710 +- 11.16467  : 1842.67730 +-  2.46465
  21000000  : 1834.41218 +- 10.92653  : 1842.28373 +-  2.40426
  22000000  : 1854.21070 +- 11.12994  : 1842.82586 +-  2.35007
 Warning: weight=-1            2       35285
  23000000  : 1841.83930 +- 10.95617  : 1842.78297 +-  2.29782
  24000000  : 1828.49319 +- 10.88035  : 1842.18756 +-  2.24825
  25000000  : 1849.61925 +- 11.05529  : 1842.48483 +-  2.20316
  26000000  : 1840.99837 +- 10.98930  : 1842.42766 +-  2.16018
 Warning: weight=-1            3       39352
  27000000  : 1839.23779 +- 10.87842  : 1842.30951 +-  2.11883
  28000000  : 1850.36220 +- 10.97517  : 1842.59711 +-  2.08042
  29000000  : 1842.48313 +- 10.89208  : 1842.59318 +-  2.04349
  30000000  : 1843.99336 +- 11.02637  : 1842.63985 +-  2.00928
  31000000  : 1846.22236 +- 10.88435  : 1842.75542 +-  1.97591
  32000000  : 1844.43879 +- 10.92402  : 1842.80802 +-  1.94436
  33000000  : 1855.61815 +- 10.91623  : 1843.19621 +-  1.91424
  34000000  : 1838.22596 +- 10.89012  : 1843.05002 +-  1.88535
  35000000  : 1838.40827 +- 10.94610  : 1842.91740 +-  1.85799
  36000000  : 1854.60202 +- 11.09578  : 1843.24197 +-  1.83249
 
 Result:
 ------- 
 Number of weighted events   =         36942025
 Average            =  1843.0071014423 fb
 Standard deviation =     1.8086070041 fb
 Maximal weight     =     0.0442221449 fb
 
 Tree-level four-fermion cross section:
 Average            =  1932.5509695232 fb
 Standard deviation =     2.9415853629 fb
 
 
 Number of events
 ----------------
 Unweighted events            =     50000
 Events with weight=-1        =         3
 Events with weight>weightmax =         1
\end{verbatim}


%
\end{document}